\documentstyle [12 pt] {article}
\title{Relativistic Green functions in a  plane wave gravitational background}
\author{A.N. Vaidya\footnote{{\it In memoriam}}, C. Farina, M.S. Guimar\~aes, M.J. Neves \\
Instituto de F\'\i sica, Universidade
Federal do Rio de Janeiro\\
21941-972, Rio de Janeiro, Brasil}         
\date{}

\begin{document}           

\maketitle                 

\hrule\hfill
\begin{abstract}
We consider a massive relativistic particle in the background of a
 gravitational plane wave. The corresponding Green functions for
 both spinless and spin 1/2 cases, previously computed by
A. Barducci and R. Giachetti \cite{Barducci3}, are reobtained here
by alternative methods, as for example, the Fock-Schwinger
proper-time method and the algebraic method.
 In analogy to the electromagnetic case, we show that for a
gravitational plane wave background a semiclassical approach is also
sufficient to provide the exact result, though the lagrangian
involved is far from being a quadratic one.
\end{abstract}
\hrule\hfill
\section{Introduction}

Green functions are basic ingredients in quantum theories.
Particularly, they are of great importance in the computation of
scattering amplitudes, as well as atomic energy levels. However,
only a few exact results for Green functions of relativistic
particles or even non-relativistic ones under the influence of
external fields are available in the literature. Of particular
importance, among the methods of obtaining Green functions is the so
called Fock-Schwinger method. It was introducted in the context of
relativistic quantum field in 1951 by Schwinger
\cite{Schwinger1951}. It has since been employed mainly in
relativistic problems such as the calculation of bosonic
\cite{Dodonov1} and fermionic \cite{Dodonov2,Dodonov3,Likken,Boschi}
Green functions in external fields. The most common exact solutions
are those involving the non-relativistic Coulomb potential
\cite{Schwinger1964}, the relativistic Dirac-Coulomb potential
\cite{Strakhovenko}, a constant and uniform electromagnetic field as
well as the electromagnetic field of a linearly polarized plane wave
\cite{Schwinger1951,ItzyksonZuber} or even particular combinations
of constant fields and plane wave fields \cite{Gitman,Dodonov1}.

In the presence of an external gravitational background the problem
becomes extremely complicated and only a few solutions using
algebraic techniques or mode summation method are known. In this
context, a simplifying assumption is to consider a weak
gravitational field. Using path integral and external source methods
it is possible to calculate the Feynman propagator in the case when
the gravitational field is that of a plane
wave\cite{Barducci1,Barducci2}. Recently, Barducci and Giachetti
\cite{Barducci3} have considered the wave equations for spin 0 and
spin ${1\over 2}$ particles in a weak external plane wave
gravitational field. They obtained for the wave functions Volkov
type solutions \cite{Volkov1935}. A similar $\it{ansatz}$ for the
Green functions is verified to be correct. These results have a
resemblance to those for a charged particle in a plane wave external
electromagnetic field. Since the latter case has been treated in a
deductive manner by the proper-time method \cite{Schwinger1951}, it
is interesting to apply this technique for the case of a weak
background of a gravitational plane wave.

The purpose of this paper is to provide many alternative methods for
computing the above mentioned Green functions. This work is
organized as follows: in section 2 we obtain the Green function for
a spin 0  massive particle in the presence of a weak background of a
gravitational plane wave by using the Fock-Schwinger method. In
section 3 we apply this same method and obtain the corresponding
Green function for a spin $1/2$ particle. Then, in section 4, we
show that  these solutions may be constructed in a much simpler
manner by an operator technique. In section 5 we reobtain the
bosonic Green function by trying a convenient ansatz. In section 6,
 we show that a path integral semiclassical aproach is sufficient
 to yield the exact results, even though the lagrangian involved
 is far from being a quadratic one. Finally, section 7 is left for
 conclusions and final remarks.

\section{Proper-time method for a scalar particle}

In the linear approximation the gravitational field is described by
a small perturbation to the flat metric, namely,
\begin{eqnarray}
g_{\mu\nu}(x)&=& \eta_{\mu\nu} + h_{\mu\nu}(x)\nonumber\\
g^{\mu\nu}(x)&=& \eta^{\mu\nu} - h^{\mu\nu}(x)
\end{eqnarray}
where $\eta_{\mu\nu}$  is the flat space metric and $h_{\mu\nu} $ is
a small perturbation. The imposition of harmonic condition in the
linear approximation gives
\begin{equation}
\partial_\mu {h^\mu}_\nu(x)={1\over 2}\partial_\nu h(x)\, .
\end{equation}
It is convenient to choose a restriction of the harmonic gauge and impose the conditions
\begin{equation}
\partial_\mu {h^\mu}_\nu = 0\, ,\;\;\;\; h(x) = 0\, .
\end{equation}
Finally, the linearized gravitational field is taken to be one
produced by a wave of arbitrary spectral composition and
polarization properties, but propagating in a fixed direction so
that
\begin{equation}
h_{\mu\nu}(x)=a_{\mu\nu} F(\xi)
\end{equation}
where $\xi=n\cdot x$ and the propagation vector is a light-like one,
which satisfies the condition $n^2=0$, and $F$ is an arbitrary
function. The conditions imposed on $h_{\mu\nu}$ give
\begin{equation}
n_\mu a^{\mu\nu} = 0\, ,\;\;\;\; Tr a = 0\, .
\end{equation}
We note that linearization implies that terms involving second and
higher powers of $a$ are to be dropped. In the following
calculations it may be convenient sometimes not to do this in the
intermediate stages. With these assumptions, the Green function for
a scalar particle in a weak external linearized gravitational field
satisfies the equation \cite{Barducci1}
\begin{equation}
\label{v01} \lbrack
{\partial^\mu}{\partial_\mu}-{h^{\mu\nu}}{\partial_\mu}{\partial_\nu}+m^2
\rbrack G(x-y) =-\delta(x-y),
\end{equation}
where $h_{\mu\nu}(x)$ is of the form described previously. In order
to apply the proper-time method we write the Green function in the
form
\begin{equation}\label{IntegralFuncaoGreen1}
G(x',x'')=-i \int_0^\infty ds \; e^{-im^2 s}\, \langle x'|
e^{-isH}|x''\rangle\, ,
\end{equation}
where the proper-time hamiltonian is given by
\begin{equation}\label{ProperTimeHParaBoson}
H= -p^2+p^{\mu}a_{\mu\nu}p^{\nu} F(\xi)\, ,
\end{equation}
with $p^\mu=i{\partial}^\mu$ and
\begin{equation}\label{CRfors=0}
[x^\mu,p^\nu]=-i\eta^{\mu\nu}.
\end{equation}

We adopt Schiwnger's notation, in which unprimed quantities are used
for operators while primed quantities are used for the corresponding
eigenvalues. In this sense, the eigenvalue equation for the operator
$x^\mu$ is written as
\begin{equation}
x^{\mu}|x'\rangle=x'^{\mu}|x'\rangle\, ,
\end{equation}
where, for simplicity, we omitted indices in the eigenvector
$|x'\rangle$. Next, given an operator $O$, we introduce the operator
$O(s)$ defined as
\begin{equation}\label{EvolucaoTemporalEms}
O(s)=e^{isH} O e^{-isH}\, ,
\end{equation}
which satisfies the Heisenberg-like equation of motion
\begin{equation}
i{d\over ds}O(s)=[O(s), H]\, .
\end{equation}
Of course, $O(0) = O$, so that $p^\mu$, $x^\mu$, etc., mean the same
as $p^\mu(0)$, and $x^\mu(0)$ and so on. Hence, we have
\begin{equation}
 x^\mu(s)\left(e^{isH}\vert x^\prime\rangle\right) =
 x' \left(e^{isH}\vert x^\prime\rangle\right)\, ,
 \end{equation}
 so that we conveniently define
 \begin{equation}
|x' s\rangle=e^{isH}|x'\rangle\, .
\end{equation}
Note that from (\ref{CRfors=0}) and (\ref{EvolucaoTemporalEms}) we
also have
\begin{equation}\label{CRfors=0}
[x^\mu(s),p^\nu(s)] = -i\eta^{\mu\nu}.
\end{equation}
Using this \lq\lq Heinsenberg picture{\rq\rq}, the Green function
can be cast into the form
\begin{equation}\label{FuncaoGreenViaSchrodingerProp}
G(x',x'') =
 -i\int_0^\infty ds \; e^{-im^2 s} \langle x' s|x'' 0\rangle \, ,
\end{equation}
where the \lq\lq Schr\"odinger-like{\rq\rq} propagator
 $\langle x's|x'' 0\rangle$ satisfies the differential equation
 \begin{equation}\label{Schrodingerlike}
 i\partial_s\langle x' s|x''0\rangle =
 \langle x' s|H|x''0\rangle\, ,
 \end{equation}
 submitted to the initial condition
 \begin{equation}\label{InitialCondition}
 \lim_{s\rightarrow 0} \langle x' s|x''0\rangle =
 \delta(x' - x'')\, .
 \end{equation}

 Schwinger's method for computing $\langle x' s|x''0\rangle$ basically consists in the following steps:
 {\it (i)} we first solve the Heisenberg equations for
 operators $p^\mu(s)$ and $x^\mu(s)$ and write the proper-time
 hamiltonian in terms of $x^\mu(s)$ and $x^\mu(0)$, instead of
 $x^\mu(0)$ $p^\mu(0)$; {\it (ii)} then, using the commutator between
$x^\mu(s)$ and $x^\mu(0)$ we write conveniently this hamiltonian in
a ordered form in time $s$, namely, with operators $x^\mu(s)$ in all
terms put on the left side, so that equation (\ref{Schrodingerlike})
can be imediately integrated in $s$ to yield
 $\langle x' s|x''0\rangle = C(x',x'')\exp\{-i\int^s F(x',x'';s')
 ds'\}$, where $F(x',x'';s') := \langle x' s|H|x''0\rangle/\langle x'
 s|x''0\rangle$; {\it (iii)} finally, the integration constant
 $C(x',x'')$ is obtained by imposing the constraints
 \begin{eqnarray}\label{Constraints}
\langle x's|p^{\mu}(s)|x''0\rangle &=& i\frac{\partial}
 {\partial x^{\,\prime}_\mu}\langle x's|x''0\rangle\, ,\cr
 \langle x's|p^{\mu}(0)|x''0 \rangle &=&
  \!\!-i\frac{\partial}{\partial x^{\,\prime\prime}_\mu}\langle x's|x''0\rangle\, ,
\end{eqnarray}
as well as the initial condition (\ref{InitialCondition}).
 A pedagogical introduction of Schwinger's method can be found in
\cite{BaroneBoschiFarinaAJP,LivroMecQuantSchwinger}(for other
applications to non-relativistic problems see references
\cite{FarinaAntonioPLA1993,HoringPRA1986})
\footnote {Schwinger's method for the non-relativistic oscillator
was developed independently by M. Goldberger and M. Gellmann in 1951
in the context of statistical mechanics \cite{Goldberger}.}
. The Heisenberg equations of motion are given by
\begin{eqnarray}\label{EquacaioParaXmu}
\frac{d}{ds} x^{\mu}(s) &=&
2\Bigl[{\eta^\mu}_{\nu}-{a^\mu}_{\nu}F\Bigl(\xi(s)\Bigr)\Bigr]p^{\nu}(s)
 \\
\frac{d}{ds}p^{\mu}(s) &=& n^{\mu} F^{\,\prime}\Bigl(\xi(s)\Bigr)
 p^{\alpha}(s)a_{\alpha\beta}p^{\beta}(s)\, ,\label{EquacaoParaPmu}
\end{eqnarray}
where in the last equation $F^{\,\prime}$ means $dF/d\xi$. In order
to solve previous equations for $x^\mu(s)$ and $p^\mu(s)$, observe
initially that they imply the following ones
\begin{eqnarray}\label{Dxids}
{d\over ds}\xi(s) &=& 2n_{\mu}p^{\mu}(s) =:2n\cdot p(s)\, ,
\\
{d\over ds}\Bigl(n\cdot p(s)\Bigr) &=& 0\, ,\label{npontop}\\
{d\over ds}\Bigl(p_{\mu}(s)a^{\mu\nu}p_{\nu}(s)\Bigr) &=& 0\,
.\label{pap}
\end{eqnarray}
Equations (\ref{Dxids}) and (\ref{npontop}) lead to
\begin{equation}
n\cdot p(s) = {{\xi(s)-\xi(0)}\over 2s}
\end{equation}
and equation (\ref{pap}) allows us to write
\begin{equation}\label{papC}
p^{\mu}(s)a_{\mu\nu}p^{\nu}(s) = C\, ,
\end{equation}
where $C$ is an operator that does not depend on $s$. In order to
integrate equation (\ref{EquacaoParaPmu}), we multiply and divide
its rhs by $d\xi/ds=2n\cdot p$ to get
\begin{equation}
\frac{d}{ds} p^\mu(s) =
 \frac{n^\mu \mbox{\Large$\frac{dF}{d\xi}$}
  \mbox{\Large$\frac{d\xi}{ds}$}\, C}
 {2n\cdot p} =
 \frac{d}{ds}\left\{ \frac{n^\mu F(\xi(s)) C}{2n\cdot p}\right\}\, ,
 \end{equation}
 where we omitted the argument $s$ in $2n\cdot p$ or $d\xi/ds$ since
 these  quantities are constants of motion and used the fact that
  $[n\cdot p,\xi]=in^2 = 0$. We also omitted the argument of $dF/d\xi$ and used that $C$ is an $s$-independent
 operator. A direct integration of the above equation leads to
\begin{equation}\label{PrimeiraIntegracaoPmu}
p^\mu(s) = \frac{n^\mu F(\xi(s))C}{2n\cdot p}\, +\, D^\mu\, ,
\end{equation}
where $D^\mu$ is a constant operator that satisfies
\begin{equation}
n_\mu D^\mu = n_\mu p^\mu = \frac{\xi(s)-\xi(0)}{2s}\, .
\end{equation}
For simplicity, from now on, we shall omit indices and a matrix
notation will be assumed, so that the previous equation is written
simply as
\begin{equation}\label{SolutionForp(s)}
p(s) = {{nF}\over 2n\cdot p}C+D\, .
\end{equation}
Two relations involving the constant operators $C$ and $D$ can be
written from our previous results. Inserting
(\ref{PrimeiraIntegracaoPmu}) into (\ref{papC}), and computing
$p^\mu p_\mu$ from last equation we obtain, respectively,
\begin{eqnarray}\label{C=DaD}
C &=& DaD\\
p^2 &=& D^2+CF\, . \label{p2=D2MaisCF^}
\end{eqnarray}
Hence, from equations  (\ref{papC}) and (\ref{p2=D2MaisCF^}), the
proper-time hamiltonian (\ref{p2=D2MaisCF^}) can be written as
\begin{equation}\label{H=-D2}
H = -D^2\, .
\end{equation}
Then, inserting (\ref{SolutionForp(s)}) into (\ref{EquacaioParaXmu})
and integrating in $s$ following a procedure analogous to that used
to obtain $p(s)$, we get
\begin{equation}
x(s)-x(0)={{A(\xi(s))-A(\xi(0))}\over 2n\cdot p}
 \left({{nC}\over n\cdot p}-2aD\right)+2Ds\, ,
\end{equation}
where we defined $A$ by
\begin{equation}
F(\xi)={{dA(\xi)}\over d\xi}\, .
\end{equation}
 Solving the previous equation for $D$, we obtain
\begin{equation}\label{D}
D=M{{x(s)-x(0)}\over 2s}-{{nC}\over 2n\cdot p}{{A(\xi(s)) -
A(\xi(0))}\over \xi(s)-\xi(0)}\; + \; {\cal O}(a^2)\, ,
\end{equation}
where
\begin{equation}
M:=\left[ I-a{{A(\xi(s))-A(\xi(0))}\over \xi(s)
-\xi(0)}\right]^{-1}\, .
\end{equation}
Equations (\ref{C=DaD}) and (\ref{D}) allow us to write constant $C$
in the form
\begin{equation}\label{Caproximado1}
C={1\over 4s^2} \Bigl(x(s)-x(0)\Bigr)a M^2\Bigl(x(s)-x(0)\Bigr)
 \; +\; {\cal O}(a^2)\, .
\end{equation}
Using equations (\ref{H=-D2}) and (\ref{D}), and keeping terms only
ut to order $a$, the proper-time hamiltonian in the weak field
approximation takes the form
\begin{equation}
H = -{1\over 4s^2}\Bigl(x(s)-x(0)\Bigr) M\Bigl(x(s)-x(0)\Bigr)\, .
\end{equation}
To write the above proper-time hamiltonian in the appropriate
s-ordered form we note the following commutation relations:
\begin{equation}
[\xi(0),x^\mu(s)]=[\xi(s)-2n\cdot p(s),x^\mu(s)]=-2in^\mu s\, ,
\end{equation}
\begin{equation}
[M^{\mu\lambda}{x_\lambda}(s),x_\mu(0)]=2 s
 \left[D^\mu+{{Cn^\mu}\over 2n\cdot p}{{A\Bigl(\xi(s)\Bigr) -
 A\Bigl(\xi(0)\Bigr)}\over \xi(s)-\xi(0)},x_\mu(0)\right] = 8is\, .
\end{equation}
Integrating equation $i\partial_s\langle x' s| x'' 0\rangle =
\langle x' s|H| x'' 0\rangle$, we get
\begin{equation}\label{PropIntegrado1}
\langle x' s|x'' 0\rangle={{ \Phi(x',x'')}\over s^2}
 \exp\left\{-{i\over 4s}(x'-x'')M(\xi',\xi'')(x'-x'')\right\}\, ,
\end{equation}
where
\begin{equation}
M(\xi',\xi'')=\left[I-a{{A(\xi')-A(\xi'')}\over \xi'-\xi''}\right]^{-1}
\end{equation}
and $\Phi(x',x'')$ is an $s$-independent quantity to be determined
by imposing constraints (\ref{Constraints}). Let us  impose the
first constraint written in (\ref{Constraints}). To evaluate its rhs
we need to compute $\partial^{\,\prime}\langle x's|x''0\rangle$,
while to evaluate its lhs we need the $s$-ordered matrix element of
operator $p(s)$. Differentiating equation (\ref{PropIntegrado1}), we
obtain
\begin{equation}
i{{\partial^{\,\prime}\langle x's|x''0\rangle}\over \langle
x's|x''0\rangle}=i{{\partial' \Phi}\over \Phi}+M{{(x'-x'')}\over
2s}+{1\over 4s}(x'-x''){\partial'M(\xi',\xi'')}(x'-x'')\, ,
\end{equation}
where
\begin{equation}
\partial^{\,\prime}M(\xi',\xi'')={n\over \xi'-\xi''}
 \left[F(\xi')-{{A(\xi')-A(\xi'')}\over \xi'-\xi''}\right]aM^2\, .
\end{equation}
Using the solution for $p(s)$, obtained from equations
(\ref{SolutionForp(s)}) and (\ref{C=DaD}), a straightforward
calculation yields
\begin{equation}
\partial^{\,\prime} \Phi(x',x'')=0\, .
\end{equation}
A similar calculation gives
\begin{equation}
\partial'' \Phi(x',x'')=0\, .
\end{equation}
Hence $\Phi(x',x'')$ is a constant, denoted simply by $\Phi$. This
constant may be evaluated by imposing the initial condition
(\ref{InitialCondition}). However, the explicit calculation is very
similar to that verifying the consistency condition in section 5, so
that we avoid it here. The result is given by
\begin{equation}
\Phi={i\over (4\pi)^2}\, .
\end{equation}
Collecting all previous results, we finally obtain
\begin{equation}
G(x',x'')={1\over 16 \pi^2}\int_0^\infty {{ds}\over s^2}\;
 e^{-im^2 s}\exp \left\{-{i\over 4s}(x'-x'')\left[I-a{{A(\xi')-A(\xi'')}\over
\xi'-\xi''}\right]^{-1}(x'-x'')\right\}\, .
\end{equation}
As expected, if we take in the above expression the limit
$a\rightarrow 0$, we recover the well known free particle propagator
in the Minkowisk space.

\section{Proper-time method  for a Dirac particle}

For a spin $1/2$ particle the Green's function satisfies the
equation
\begin{equation}
\left(i\gamma^\mu\partial^{\,\prime}_\mu-{i\over
2}F(\xi^{\,\prime})a^{\mu\nu}\gamma_\mu\partial^{\,\prime}_\nu -
 m\right)S_F(x^{\,\prime},x^{\,\prime\prime}) =
 \delta(x^{\,\prime} - x^{\,\prime\prime})\, ,
\end{equation}
where $\gamma^\mu$ are the usual Dirac matrices. As it is common in
problems involving fermion Green functions in external fields, we
conveniently define $\Delta_F(x^{\,\prime},x^{\,\prime\prime})$ as
follows:
\begin{equation}\label{SFuncaodeDelta}
S_F(x^{\,\prime},x^{\,\prime\prime}):=
\left(i\gamma^\mu\partial^{\,\prime}_\mu-
 {i\over
 2}F(\xi^{\,\prime})a^{\mu\nu}\gamma_\mu\partial^{\,\prime}_\nu +
  m\right)\Delta_F(x^{\,\prime},x^{\,\prime\prime})\, .
\end{equation}
As a consequence, $\Delta_F(x,y)$  satisfies a boson-like second
order differential equation, namely,
\begin{equation}
\left(\partial^{\,\prime\mu}\partial^{\,\prime}_\mu
 -  F(\xi^{\,\prime}) a^{\mu\nu}\partial^{\,\prime}_\mu\partial^{\,\prime}_\nu
 + {i\over 2}{{dF}\over d\xi^{\,\prime}}
  \,\sigma_{\mu\nu}n^{\mu}a^{\nu\rho}
 \partial^{\,\prime}_\rho + m^2\right)\Delta_F(x^{\,\prime},x^{\,\prime\prime}) =
  -\delta(x^{\,\prime} - x^{\,\prime\prime})\, ,
\end{equation}
 where $\sigma_{\mu\nu}=(i/2)[\gamma_\mu,\gamma_\nu]$ and we neglected terms
 quadratic in $a$. Using the
 notation introduced previously, $\Delta_F(x',x'')$
 can be written as
\begin{equation}
\Delta_F(x',x'') =
 -i\int_0^\infty \!\! ds \, e^{-im^2 s} \langle x'|e^{-iHs} |x''\rangle =
 -i\int_0^\infty \!\! ds\, e^{-im^2 s} \langle x's|x''0\rangle
 \, ,
\end{equation}
where now the corresponding proper-time hamiltonian can be written
in the form
\begin{equation}\label{ProperTimeHParaFermion}
H= -p^2+pap F(\xi)+{1\over 2}pa\sigma n{{dF(\xi)}\over d\xi}\, ,
\end{equation}
where we are following the same notation as before, namely, primed
quantities are eigenvalues while unprimed ones are operators. In
writing last equation we also used the fact that $[pap,F(\xi)] = 0$
and $[pa\sigma n,dF(\xi)/d\xi]=0$.

From now on, whenever it does not cause any confusion, we shall omit
the argument $s$ from the operators involved as well as the argument
of $F$ and its derivatives $dF/d\xi$ and $d^2F/d\xi^2$. With this in
mind, the Heisenberg equations of motion for the operators $x$ and
$p$ are given, respectively, by
\begin{eqnarray}\label{HEforX}
{dx\over ds}&=&2(I-aF)p-{1\over 2}a\sigma n {{dF}\over d\xi}\, ,\\
{dp\over ds}&=&n\left(pap{{dF}\over d\xi} +
 {1\over 2}pa\sigma n{{d^2 F}\over d\xi^2}\right)\, .\label{HEforP}
\end{eqnarray}
In this case, the constants of motion are
\begin{eqnarray}\label{nPontopFermion}
n\cdot p&=&{{\xi(s)-\xi(0)}\over 2s}\, ,\\
C_1 &=& pap\, ,\label{C1Fermion}\\
C_2 &=& pa\sigma n\, .\label{C2Fermion}
\end{eqnarray}
Integrating the equation of motion for $p$ we have
\begin{equation}\label{SolucaoP(s)ParaFermion}
p(s)={n\over 2n\cdot p}\left(C_1F+{1\over 2}C_2 {{dF}\over
d\xi}\right)+D_f\, ,
\end{equation}
where $D_f$ is a constant. For future convenience, we use the
previous equation to write $p^2$ in the form
\begin{equation}\label{p2ParaFermion}
p^2 = D_f^2 + C_1 F +\frac{1}{2} C_2 \frac{dF}{d\xi}\, ,
\end{equation}
where we used the fact that $n\cdot p = n\cdot D_f$. Substituting
equation (\ref{SolucaoP(s)ParaFermion}) into the Heisenberg equation
(\ref{HEforX}) and integrating in $s$, we obtain, after some
convenient rearrangement,
\begin{eqnarray}\label{DParaFermion}
D_f &=& \left[I-a{{A\Bigl(\xi(s)\Bigr) - A\Bigl(\xi(0)\Bigr)}\over
\xi(s)-\xi(0)}\right]^{-1} \Biggl\{{{x(s)-x(0)}\over 2s}\Biggr\}
 + {1\over 4}a\sigma n{{F(\xi(s))-F(\xi(0))}\over \xi(s)-\xi(0)}
 \nonumber\\
 &-&
  {n\over 2n\cdot p}\left[C_1{{A(\xi(s))-A(\xi(0))}\over
\xi(s)-\xi(0)}+{1\over2}C_2{{F(\xi(s))-F(\xi(0))}\over
\xi(s)-\xi(0)}\right] + {\cal O}(a^2) \, .
\end{eqnarray}
Since $n^\mu a_{\mu\nu} = 0$, and keeping only terms up to first
order in $a$, constants $C_1$ and $C_2$ are given, respectively, by
\begin{eqnarray}
C_1 &=& D_faD_f = \left({{x(s)-x(0)}\over 2s}\right) a
 \left({x(s)-x(0) \over 2s}\right) + {\cal O}(a^2)\\
C_2 &=& D_fa\sigma n =  \left({{x(s)-x(0)}\over 2s}\right) a\sigma n
+ {\cal O}(a^2)\, .
\end{eqnarray}
From equations (\ref{ProperTimeHParaFermion}) and
(\ref{p2ParaFermion}), the proper-time hamiltonian for the case of a
Dirac particle is given by
\begin{equation}
H = -D_f^2\, ,
\end{equation}
which, in the linear approximation (weak gravitational field), can
be written as
\begin{equation}
H = -\left({{x(s)-x(0)}\over 2s}\right)
\left[I+a{{A(\xi(s))-A(\xi(0))}\over \xi(s)-\xi(0)}\right]
 \left({{x(s)-x(0)}\over 2s}\right)\, .
\end{equation}
One may repeat the steps followed in the last section to get
\begin{equation}
\langle x' s|x'' 0 \rangle={{\Phi(x',x'')}\over s^2}
 \exp \left\{-{i\over 4s}(x'-x'')\left[I+a{{A(\xi')-A(\xi'')}\over
\xi'-\xi''}\right](x'-x'')\right\}\, ,
\end{equation}
where $\Phi(x',x'')$ is to be determined. The constraints to be
satisfied are the same as in the previous case, namely,
$ \langle x's|p(s)|x''0\rangle=i\partial^{\,\prime}\langle
x's|x''0\rangle$ and
 $\langle x's|p(0)|x''0\rangle=-i\partial''\langle
x's|x''0\rangle$. The first constraint leads to
\begin{equation}
\partial^{\,\prime} \Phi =
\left[ \partial^{\,\prime} {\langle{C_2}\rangle\over 4 n\cdot
p}(F(\xi')-F(\xi''))\right]\Phi\, ,
\end{equation}
where
\begin{equation}
 \langle C_2\rangle := \langle x' s|C_2|x''0\rangle = {{(x'-x'')}\over
2s}a\sigma n \, + \, {\cal O}(a^2)\, .
\end{equation}
Hence, we can write
\begin{equation}
\Phi(x',x'')=\chi(x'')\exp\left\{{{s\langle C_2\rangle
(F(\xi')-F(\xi''))}\over 2(\xi'-\xi'')}\right\}\, ,
\end{equation}
where the $x''$-dependence of $\chi(x'')$ is determined by the
second constraint written previously. The imposition of this second
constraint leads to the differential equatin $\partial''\chi = 0$,
so that $\chi$ is a constant, denoted by $C_0$:
\begin{equation}
\Phi(x',x'')= C_0\exp\left\{{{s\langle C_2\rangle
(F(\xi')-F(\xi''))}\over 2(\xi'-\xi'')}\right\}\, .
\end{equation}
As before, we evaluate the remaining constant $C_0$ by taking the
limit $s\to 0$ and the final result for $\Delta_F(x',x'')$ is
written as
\begin{eqnarray}
\Delta_F(x',x'')&=&{1\over 16 \pi^2}\int_0^\infty {{ds}\over s^2}\; e^{-im^2 s}\exp \left\{-{i\over 4s}(x'-x'')\left[I+a{{A(\xi')-A(\xi'')}\over \xi'-\xi''}\right](x'-x'')\right\}
\nonumber\\
&\times&\exp \left\{{(x'-x'')a\sigma n (F(\xi')-F(\xi''))}\over
4(\xi'-\xi'')\right\}\, .
\end{eqnarray}
The desired fermionic Green function is then obtained by inserting
last expression into equation (\ref{SFuncaodeDelta}).

\section{Green functions by operator techniques}

In this section we show that using operator techniques one can
obtain the Green functions in a simple manner. We were motivated by
successful calculations previously made for Green functions of
relativistic charged particles in similar external electromagnetic
fields \cite{Boschi,VaidyaJPA1988}. We shall start by discussing the case of
a scalar particle. The essential idea is to start with  expression
(\ref{FuncaoGreenViaSchrodingerProp}) for the Green function, namely,
$G(x',x'')=-i\int_0^\infty ds \; e^{-im^2 s} \langle x' s|x''
0\rangle$ and then relate by an algebraic method the expression of
Schrödinger-like propagator $\langle x' s| x'' 0\rangle$ for the problem containing the
interaction with the corresponding one for the free case, denoted by
$\langle x' s| x'' 0\rangle_0$. For the scalar particle  we note that
\begin{equation}
p^2-pap F(\xi)= S_0 p^2 {S_0}^{-1}\, ,
\end{equation}
where we defined the operator $S_0$ by
\begin{equation}\label{S0}
S_0=exp\left[-i{{papA(\xi)}\over 2n\cdot p}\right]
\end{equation}
and used the well known relation for operators $A$ and $B$
\begin{equation}\label{BCH}
e^A B e^{-A} = \sum_n C_n\; ,\;\;\;\mbox{where}\;\;\;\;
C_0=B\;\;\;\;\mbox{and}\;\;\;\; C_{n+1} = [A,C_n]\, .
\end{equation}
As a consequence, we have
\begin{eqnarray}\label{FatoracaoAlbebricaComS0}
\langle x' |e^{is(p^2-pap F)}|x''\rangle &=&
 \langle x'|S_0e^{isp^2}{S_0}^{-1}|x''\rangle\cr
 &=&
 \langle x'|e^{isp^2}\left(e^{-isp^2} S_0 e^{isp^2}\right){S_0}^{-1}|x''\rangle\cr
 &=&
 \langle x's|S_0(s) {S_0(0)}^{-1}|x''0\rangle_0\, ,
\end{eqnarray}
where we conveniently inserted the identity operator
$e^{isp^2}e^{-isp^2}$ and the meaning of the subscript
{\footnotesize o} is such that
\begin{equation}
\langle x's|x''0\rangle_0={i\over (4\pi s)^2}e^{-i{{(x'-x'')^2}\over
4s}}\; .
\end{equation}
Inserting (\ref{S0}) into (\ref{FatoracaoAlbebricaComS0}), a
straightforward calculation leads to the matrix element
\begin{equation}
\langle x's|x''0\rangle={1\over (4\pi s)^2}
 \exp \left\{-{i\over 4s}(x'-x'')
 \left[I+a{{A(\xi')-A(\xi'')}\over
 \xi'-\xi''}\right](x'-x'')\right\}\, .
\end{equation}
as an exact result. It is also the same as that found by the proper-time method in the small $a$ approximation. This is curious. Perhaps the reason is that we have imposed very few restrictions on $a$. In the case of the electromagnetic field the tensor $f_{\mu\nu}$ and its dual have special properties which allow simplifications. The other difference is that the proper-time equations of motion are nonlinear.

For the Dirac particle we may proceed in an analogous way. With this purpose, we now define
operator $S$ as
\begin{equation}
S=exp\left[{i\over 4}{{pa\sigma n}\over n\cdot p}F(\xi)\right]\, .
\end{equation}
Using the BCH-like relation (\ref{BCH}) it is straightforward to show that
\begin{equation}
p^2-pap F(\xi)-{i\over 2}pa\sigma n{{dF}\over d\xi}
=S(p^2-pap F(\xi))S^{-1}\, .
\end{equation}
Note that operator $S$ eliminates the interaction term containing explicitly the Dirac matrices.
To reduce further the remaining term can be done with the aid of operator $S_0$, introduced
before for the spinless particle. Hence, we obtain for the Dirac case, the analogous expression of equation
(\ref{FatoracaoAlbebricaComS0}), namely,
\begin{equation}
\langle x'|exp\left\{is\left[p^2-pap F(\xi) -
{i\over 2}pa\sigma n{{dF}\over d\xi}\right]\right\}|x''\rangle =
\langle x's|S(s)S_0(s)(S(0)S_0(0))^{-1}|x''0\rangle_0\, .
\end{equation}
Inserting in the previous equation the expressions of $S(s)$, $S_0(s)$, $S_0(0)^{-1}$ and
$S(0)^{-1}$ it is not difficult to obtain the desired Green function for the Dirac particle.

We close this section by showing an alternative way of employing albegraic methods
in order to obtain the above relativistic Green functions. As we shall see, the
present method leads directly to the Green functions without using the proper-time method.

Starting with the spin $0$ case, we have
\begin{equation}
G(x',x'')=\langle x'|G|x''\rangle\, ,
\end{equation}
where
\begin{equation}
G = \left(p^2-pap F(\xi)-m^2\right)^{-1} = S_0(p^2-m^2
)^{-1}{S_0}^{-1}\, ,
\end{equation}
with operator $S_0$ defined by equation (\ref{S0}). Hence, we can write
\begin{equation}
G_F(x',x'') = S_0(i\partial^{\,\prime})
\int {{{d^4}p}\over (2\pi)^4}{{e^{-ip\cdot(x'-x'')}}\over p^2-m^2+i\epsilon}
  {S_0}^{-1}(-i\partial'')
\end{equation}
which leads to the final result
\begin{equation}
G_F(x',x'') = \int {{{d^4}p}\over (2\pi)^4}
 {{e^{-ip\cdot(x'-x'')}}\over p^2-m^2+i\epsilon} exp\left\{-i{{pap
}\over 2n\cdot p}[A(\xi')-A(\xi'')]\right\}
\end{equation}

For the spin ${1\over 2}$ the Green function $S_F(x',x'') $ is given by
\begin{equation}
S_F(x',x'') = \left( i\gamma^\mu \partial^{\,\prime}_\mu -
\frac{i}{2} F(\xi^\prime) a^{\mu\nu}\gamma_\mu\partial^{\,\prime}_\nu + m\right)
\Delta_F(x',x'')\, ,
\end{equation}
where $\Delta_F(x',x'')$ is given by
\begin{equation}
\Delta_F(x',x'')=\int {{{d^4}p}\over (2\pi)^4}
 {{e^{-ip\cdot(x'-x'')}}\over p^2-m^2+i\epsilon}
  exp\left\{-i{{pap }\over 2n\cdot p}[A(\xi')-A(\xi'')]\right\}
  exp\left\{{i\over 4}{{pa\sigma n}\over n\cdot p}[F(x')-F(x'')]\right\},
\end{equation}
which agrees with the result obtained in section {\bf 2.2}.
%


\section{Ansatz solution for the Green function}

In this section we presen an alternative procedure to construct
previous Green functions which is based on a convenient ansatz for
the desired solution. We ilustrate the method for the spin 0
particle but a generalization to the  spin 1/2 case can be made
without difficulty. The main motivation for such a procedure is that
it worked very well for the caso of a relativistic charged particle
under  an external electromagnetic field of a plane wave
\cite{VaidyaNuovoCimento1988}. The desired Green function satisfies
the differential equation
\begin{equation}\label{e03}
\left[\partial^\mu\partial_ \mu -
a_{\mu\nu}F(\xi)\partial^{\mu}\partial^{\nu} +
 m^2\right]G(x-x')=-\delta(x-x')\, ,
\end{equation}
where, for convenience, we are using in this section variables $x$
and $x'$, instead of variables $x'$ and $x''$ chosen in the
preceding section. Hence, along this section, $x$ is not an
operator. As usual, the Green function can be written as
\begin{equation}\label{GDelta}
G(x,x')=\int^{\infty}_{0}ds \; e^{-im^2s}\Delta(x,x',s)\, ,
\end{equation}
where $\Delta(x,x',s)$ satisfies the Schrödinger-like differential equation
\begin{equation}\label{e05}
\left[i\partial_{s} - \partial^{\mu}\partial_{\mu} +
 a_{\mu\nu}F(\xi)\partial^{\mu}\partial^{\nu}\right]\Delta(x,x',s)=0,
\end{equation}
subjected to the initial condition
\begin{equation}
i\Delta(x,x',s)\stackrel{s\rightarrow 0^{+}}{\longrightarrow}\delta(x-x')\, .
\end{equation}
Now, in order to factorize the free particle solution, we try the following ansatz
\begin{equation}\label{e07}
\Delta(x,x',s) = \Delta_{0}(x,x',s)\Sigma(x,x',s)\, ,
\end{equation}
where $\Delta_{0}(x,x')$ corresponds to the free particle solution, that is,
\begin{equation}\label{e11}
\left(i\partial_{s} - \partial^2\right)\Delta_{0}(x,x') = 0\, ,
\end{equation}
whose the well known solution is given by
\begin{equation}
\Delta_{0}(x,x',s) = \frac{i}{16\pi^2
s^2}\exp\left\{-i\frac{(x-x')^2}{4s}\right\}\, .
\end{equation}
Substituing  our ansatz (\ref{e07}) into equation  (\ref{e05}), we have
\begin{eqnarray}
\label{e10}
  \Delta_{0}i\partial_{s}\Sigma -
  2 \partial_{\mu}\Delta_{0}\partial^{\mu}\Sigma
  \!\!\!\! &-& \!\!\!\!\Delta_{0}\partial^2\Sigma +
  Fa_{\mu\nu}\left(\partial^{\mu}\partial^{\nu}\Delta_{0}\right)\Sigma +\nonumber\\
&+& \!\!2Fa_{\mu\nu}\partial^{\mu}\Delta_{0}\partial^{\nu}\Sigma \;+
 Fa_{\mu\nu}\left(\partial^{\mu}\partial^{\nu}\Sigma\right)\Delta_{0} \; = 0.
\end{eqnarray}
Since  $h_{\mu\nu}$ is small, we neglect the last two terms in above and get
\begin{equation}\label{EqParaSigma}
  \Delta_{0}i\partial_{s}\Sigma - 2 \partial_{\mu}\Delta_{0}\partial^{\mu}\Sigma -
  \Delta_{0}\partial^2\Sigma +
  Fa_{\mu\nu}\left(\partial^{\mu}\partial^{\nu}\Delta_{0}\right)\Sigma = 0.
\end{equation}
Next, using that
\begin{eqnarray}
\partial_\mu \Delta_0 &=& -i{{(x-x')_\mu}\over 2s}\Delta_0\nonumber\\
\partial_\mu\partial_\nu\Delta_0
 &=&
 -i{{\eta_{\mu\nu}}\over 2s}\Delta_0-{{(x-x')_\mu (x-x')_\nu}\over 4s^2}\Delta_0
\end{eqnarray}
equation (\ref{EqParaSigma}) takes the form
\begin{equation}\label{EqParaSigma2}
\left[i\partial_{s}-\partial_\mu\partial^{\mu}+i{{(x-x')_\mu}\over s}\partial^{\mu} -
{F\over 4s^2}(x-x')_\mu a^{\mu\nu} (x-x')_\nu\right] \Sigma = 0\, .
\end{equation}
An inspection of  the previous equation suggests us to try a solution of the form
\begin{equation}
\Sigma=exp\left\{-i{f(\xi,\xi')(x-x')_\mu a^{\mu\nu} (x-x')_\nu}\over 4s\right\}\, .
\end{equation}
%
%
In order to substitute  last expression into equation (\ref{EqParaSigma2}), we need
\begin{equation}\label{DsSigma}
i\partial_s \Sigma = -f{{(x-x')_\mu a^{\mu\nu} (x-x')_\nu}\over 4s^2}\Sigma\, ,
\end{equation}
and
\begin{equation}\label{D1Sigma}
\partial_\mu \Sigma =
-{i\over 4s}\left[n_\mu{df\over d\xi}{(x-x')_\mu a^{\mu\nu} (x-x')_\nu} +
 2fa_{\mu\nu}(x-x')^\nu \right] \Sigma\, .
\end{equation}
Further, we also have
\begin{eqnarray}
\partial_\mu\partial^\mu\Sigma
 = &-&
 \!\!\! {1\over 16s^2}
 \left[n_\mu{df\over d\xi}{(x-x')_\alpha a^{\alpha\beta} (x-x')_\beta} +
  2fa_{\mu\alpha}(x-x')^\alpha \right] \times\nonumber\\
&\times&
 \!\!\! \left[n^\mu{df\over d\xi}{(x-x')_\alpha a^{\alpha\beta}
(x-x')_\beta} + 2fa^{\mu\alpha}(x-x')_\alpha\right]\Sigma\, ,
\end{eqnarray}
where we have used the properties of $n_\mu$ and $a_{\mu\nu}$,
namely, $n^2=0$ and $n_\mu a^{\mu\nu} = 0$. To the order of
magnitude of our interest we may write
\begin{equation}\label{D2Sigma}
\partial_\mu\partial^\mu\Sigma=0
\end{equation}
Hence, substituing equations (\ref{D2Sigma}), (\ref{D1Sigma}) and (\ref{DsSigma})
into equation (\ref{EqParaSigma2}), we get
\begin{equation}\label{e16}
 \frac{d}{d\xi}\left[(\xi-\xi')f\right] = F
\end{equation}
 We have then
\begin{equation}
f(\xi,\xi')= \frac{A(\xi)-A(\xi')}{\xi-\xi'}.
\end{equation}
where we defined function $A$ by
\begin{equation}
F = \frac{dA}{d\xi}\, .
\end{equation}
Therefore, equation (\ref{e07}) takes the form
\begin{eqnarray}\label{e17}
\Delta(x,x',s) &=& \frac{i}{16\pi^2 s^2}
 \exp\left\{-i\frac{(x-x')^2 +
 (x-x')^{\mu}a_{\mu\nu}\Omega(\xi)(x-x')^{\nu}}{4s}\right\}\nonumber\\
 &=& \frac{i}{16\pi^2 s^2}
 \exp\left\{-i\frac{(x-x')^2+(x-x')^{\mu}a_{\mu\nu}\frac{A(\xi) -
 A(\xi')}{\xi-\xi'}(x-x')^{\nu}}{4s}\right\}\, .
\end{eqnarray}
With the purpose of checking the self-consistency of the result just obtained, not that
\begin{equation}
\lim_{s\rightarrow 0^+}\int d^4k \;
 e^{i{k^{\mu}(B_{\mu\nu}k^{\nu}s}-\Delta x_{\mu})}=\lim_{s\rightarrow
0^+}\exp\left\{-i\frac{\Delta x_{\mu} \left(B^{-1}\right)^{\mu\nu}
\Delta x_{\nu}}{4s}\right\}\frac{i(\pi)^2}{s^2\sqrt{det(B)}}\, .
\end{equation}
As a consequence, we have
\begin{equation}
\lim_{s\rightarrow 0^+}\frac{i}{16\pi^2 s^2}
 \exp\left\{-i\frac{\Delta x_{\mu}
  \left(B^{-1}\right)^{\mu\nu} \Delta x_{\nu}}{4s}\right\} =
 {\delta^{4}}(x-x')\sqrt{det(B)}\, .
\end{equation}
Making the following identifications,
\begin{eqnarray}
({B^{\mu}}_{\nu})^{-1}={\delta^{\mu}}_{\nu} +
{a^{\mu}}_{\nu}\frac{\Delta A}{\Delta \xi} \hspace{2cm} \nonumber \\
{B^{\mu}}_{\nu} = {\delta^{\mu}}_{\nu} -
{a^{\mu}}_{\nu}\frac{\Delta A}{\Delta \xi} \hspace{2cm} \nonumber
\end{eqnarray}
we get
\begin{equation}
det(B)=1+Tr({a^{\mu}}_{\nu})\frac{ A(\xi)-A(\xi')}{ \xi-\xi'} = 1\,
,
\end{equation}
which confirms the  initial condition in the parameter $s$,
\begin{equation}\label{e06}
i\Delta(x,x',s)\stackrel{s\rightarrow 0^{+}}{\longrightarrow}\delta(x-x')\, .
\end{equation}
Inserting expression (\ref{e17}) into equation (\ref{GDelta}), we obtain the
desired Green function, in agreement with our previous calculations.

\section{Semiclassical approximation}

As we have seen in many of the previous calculations, relativistic
Green functions may be written in terms of Schr\"odinger-like
propagators, if we introduce appropriately an integration over the
so called proper-time $s$. For instance, the Green function for a
scalar particle can be written as $G_F(x',x'') = -i\int_0^\infty \,
ds\, e^{-is(m^2-i\epsilon)} \langle x' s|x'' 0\rangle$ (see equation
(\ref{IntegralFuncaoGreen1})), where $\langle x' s|x'' 0\rangle$ can
be interpreted as a Feynman propagator of an auxiliary problem of a
non-relativistic particle in four dimensions whose dynamics
corresponds to the evolution in the parameter $s$, which plays the
role of time in this auxiliary problem. Once $\langle x' s|x''
0\rangle$ behaves like a non-relativistic Feynman propagator, we
have at our dispposal all techniques developed to compute this
quantity as, for example, the Feynman path integral method. In
particular, in the context of path integrals, it may be convenient
to use the semiclassical approximation. It is well known that
whenever the corresponding lagrangian is quadratic in the
coordinates and velocities the semiclassical result gives the exact
result.

The purpose of this section is to show that if we apply the
semiclassical approximation to the problem at hand (a relativistic
particle in the background of a plane wave gravitational field) we
shall obtain the exact result. At first sight this is an unexpected
result, since the lagrangian of the corresponding classical problem
is far from being quadratic. However, there is a strong reason which
suggests that this will be the case, namely, the semiclassical
method when applied to the problem of a relativistic charged
particle in an external field of an electromagnetic plane wave
yields the exact result. Here, we shall discuss only the case of a
scalar particle. To avoid any confusion with the notation, observe
that, in this section, all quantities are not operators, but
classical numbers.

The semiclassical approximation for the Feynman propagator $\langle
x' s|x'' 0\rangle$ is given by
\begin{equation}\label{SemiClassicalPropagator}
\langle x' s|x'' 0\rangle=\frac{1}{(2\pi i)^2}\;
\left|\frac{\partial^{2}S_{cl}}{\partial x'\partial
x''}\right|^{1/2}\;e^{iS_{cl}}
\end{equation}
where $S_{cl}$ means the functional action
\begin{equation}
S(x_\mu) =
 \int_0^{s} L\Bigl( x_\mu(\tau), {\dot x}_\mu(\tau)\Bigr)\, d\tau
\end{equation}
evaluated with the classical solution $x_{cl}^\mu$, that is,
 $S_{cl} = S(x_{cl})$, where $x_{cl}$ satisfies the Euler-Lagrange equations
\begin{equation}\label{EulerLagrange}
\frac{d}{ds}\left.\left(\frac{\partial L}{\partial {\dot
x}_\mu}\right)\right|_{x^\mu=x_{cl}^\mu} = \left.\frac{\partial
L}{\partial x_\mu}\right|_{x^\mu=x_{cl}^\mu} \, , \;\;\;\;\;\;\mu =
0,1,2,3\, ,
\end{equation}
submitted to the Feynman conditions
\begin{equation}\label{FeynmanConditions}
x_{cl}^\mu(\tau=0) = x^{\,\prime\prime\mu}\,\, ;\;\;\;\;
x_{cl}^\mu(\tau=s) = x^{\,\prime\mu}\, .
\end{equation}
Hence, in order to use this approximation, we need to construct the
classical lagrangian corresponding to the following classical
hamiltonian
\begin{equation}
H(x,p)=-p^2+p^{\mu}h_{\mu\nu}(x)p^{\nu}\, .
\end{equation}
Recalling that $p_\mu = \partial L/\partial{\dot x}^\mu$, it is not
difficult to show that the corresponding lagrangian can be written
as
\begin{equation}\label{Lagrangian}
L(x,\dot{x}) = -\frac{1}{4}\Bigl(\dot{x}_{\mu}\dot{x}^{\mu} +
\dot{x}^{\mu}h_{\mu\nu}(x)\dot{x}^{\nu}\Bigr) =
 -\frac{1}{4}g_{\mu\nu}(x)\dot{x}^{\mu}\dot{x}^{\nu}\, .
\end{equation}
The functional action is then given by
\begin{equation}\label{DefinitionFunctionalAction}
S(x) = \int_{0}^{s}
\left[-\frac{1}{4}\dot{x}^{\mu}(\tau)\dot{x}_{\mu}(\tau) -
\frac{1}{4}\dot{x}^{\mu}(\tau)a_{\mu\nu}\dot{x}^{\nu}(\tau)
 F(n\cdot x(\tau))\right]\, d\tau \;.
\end{equation}

The application of Euler-Lagrange equations (\ref{EulerLagrange}) to
the lagrangian (\ref{Lagrangian}) lead to classical equations
completely analogous to the Heisenberg equations discussed in
section 2, but do not forget that here $x_\mu$ and $p_\mu$ ara not
operators. However, exactly the same kind of solutions are obtained
from these equations, so that we just write the classical solution
as
\begin{equation}\label{sc01}
 x_{cl}^{\mu}(\tau) = x_{cl}^{\mu}(0)+\left[\frac{n^{\mu}C}{2(n\cdot p)^2} -
\frac{a^{\mu\nu}D_{\nu}}{n\cdot
p}\right]\left[A(\xi(\tau))-A(\xi(0)) \right]-2D^{\mu}\tau\, ,
\end{equation}
where $C$ and $D_{\mu}$ are (classical) constants to be determined
by imposing Feynman conditions (\ref{FeynmanConditions}). From the
classical version of (\ref{Caproximado1}) we have, up to first order
in $a$,
\begin{equation}\label{Cultima}
C = \frac{1}{4}\,\frac{(x' -x'')_\mu a^{\mu\nu}(x' -x'')_\nu}{s^{2}}
=: \frac{1}{4}\,\frac{\Delta x^\mu a_{\mu\nu} \Delta x^\nu}{s^2}\, ,
\end{equation}
where we defined $\Delta x^\mu = (x' -x'')^\mu$.

In order to obtain $D^\mu$, we take $\tau = s$ in  equation
(\ref{sc01}) and contract $x_{cl}^{\mu}(s)$ with $n_{\mu}$ and
$a^{\mu\nu}$ to obtain, respectively,
\begin{eqnarray}\label{sc02}
n^{\mu}\Delta x_{\mu}&=&-2D^{\mu}n_{\mu}s
 \;\;\;\;\;\Longrightarrow\;\;\;
 n\cdot D = -\frac{\Delta \xi}{2s}\, ;\\
a^{\mu\nu}\Delta x_{\nu}&=&-2D_{\nu}a^{\mu\nu}s\, ,
  \;\;\;\Longrightarrow\;\;\;
 a^{\mu\nu}D_\nu = -\frac{a^{\mu\nu}\Delta x_ \nu}{2s}\, ,
\end{eqnarray}
where we defined $\Delta \xi = n^\mu(x' -x'')_\mu = \xi^{\,\prime}
 -\xi^{\,\prime\prime}$. Substituting these relations
 into the solution (\ref{sc01}) with $\tau = s$, we obtain $D^{\mu}$
 in terms of $x'$ and $x''$, namely,
\begin{eqnarray}\label{DmuUltima}
D^{\mu}=\frac{1}{2s}
 \left(n^{\mu}\frac{\Delta
x_{\alpha}a^{\alpha\beta}\Delta x_{\beta}}{2(\Delta\xi)^2} +
\frac{a^{\mu\nu}\Delta x_{\nu}}{\Delta\xi} \right)\Delta A
-\frac{\Delta x^{\mu}}{2s}\, ,
\end{eqnarray}
where we $\Delta A:= A(\xi^{\,\prime}) - A(\xi^{\,\prime\prime})$.
Substituting expressions (\ref{Cultima}) and (\ref{DmuUltima}) into
equation (\ref{sc01}), and then differentiating with respect to
$\tau$, we obtain the classical velocity ${\dot x}_{cl}^\mu (\tau)$
at any \lq\lq instant{\rq\rq} $\tau$ in terms of $x'$ and $x''$,
namely,
\begin{eqnarray}
\dot{x}^{\mu}(\tau)=\left(n^{\mu}\frac{\Delta
x_{\alpha}a^{\alpha\beta}\Delta x_{\beta}}{2(\Delta\xi)^2} +
\frac{a^{\mu\nu}\Delta
x_{\nu}}{\Delta\xi}\right)\frac{dA}{d\tau}-2D^{\mu}\, .
\end{eqnarray}
Substituting last equation into the functional action
(\ref{DefinitionFunctionalAction}) and keeping only terms ur to
first order in $a$, we obtain after a lengthy but straightforward
 calculation the desired classical action,
\begin{eqnarray}
S_{cl} &=&
 - \frac{1}{4s}\Delta
x_{\mu}\left(\eta^{\mu\nu}+a^{\mu\nu}\frac{\Delta A}{\Delta\xi}
\right)\Delta x_{\nu}  \nonumber\cr\cr
 &=&
 - \frac{1}{4s} \Bigl( x' - x''\Bigr)^\mu
 \left( \eta_{\mu\nu} + a_{\mu\nu}\,
 \frac{A(\xi') - A(\xi'')}{\xi' - \xi''}\right)
 \Bigl( x' - x''\Bigr)^\nu, .
\end{eqnarray}
Now we need to evaluate the Van Vleck-Pauli-Morette determinant to
obtain the pre-exponencial factor of the semiclassical propagator.
With this purpose, first note that
\begin{equation}
\frac{\partial^{2}S_{cl}}{\partial {x'}_\mu\partial {x''}_\nu} =
 \frac{1}{2s}\Bigl(\eta^{\mu\nu}+{\cal O}(a)\Bigr) \, .
\end{equation}
As a consequence, we have
\begin{equation}
\det\left(\frac{\partial^{2}S_{cl}}{\partial x'\partial x''}\right)
 =
 \frac{1}{16s^4}\det\Bigl(\eta + {\cal O}(a)\Bigr)
 =
 \frac{1}{16s^4}\Bigl(1+ \mbox{Tr}{\cal O}(a) \Bigr)\, .
\end{equation}
However, note that $tr({\cal O}(a)) = {\cal O}(a^2)$ so that
$\det^{1/2}\left(\frac{\partial^{2}S_{cl}}{\partial x'\partial
x''}\right)=\frac{1}{4s^2} + {\cal O}(a^2)$. Collecting the previous
results and using equation (\ref{SemiClassicalPropagator}) we
conclude that, up to first order in $a$, the Feynman propagator
$\langle x' s|x'' 0\rangle$ in the semiclassical aproximation is
given by
\begin{eqnarray}
\langle x' s|x'' 0\rangle \!\! &=& \!\!\!\!\frac{1}{16\pi^2s^2}
\exp\!\left\{\frac{-i}{4s}\Delta
x_{\mu}\left(\eta^{\mu\nu}+a^{\mu\nu}\frac{\Delta A}{\Delta\xi}
\right)\Delta x_{\nu}\right\}\cr\cr
 &=&
 \!\!\!\!\frac{1}{16\pi^2s^2}  \exp\!\left\{\frac{-i}{4s}
 \Bigl( x' - x''\Bigr)^\mu
 \left(\eta_{\mu\nu} + a_{\mu\nu}
 \frac{A(\xi') - A(\xi'')}{\xi' - \xi''} \right)
 \Bigl( x' - x''\Bigr)^\nu\right\}.
\end{eqnarray}
Substituting last expression into equation
(\ref{FuncaoGreenViaSchrodingerProp}), we reobtain the correct Green
function for a scalar relativistic particle in the weak
gravitational field of a plane wave.


\section{Conclusions and final remarks}

In this work we have presented a few alternative techniques to
calculate Green functions of relativistic particles  under the
influence of a weak gravitational field of a plane wave, with
particular attention to the Fock-Schwinger proper-time method. In
fact, we started this paper by a detailed application of this method
in the construction of the Green functions for both spin zero as
well as spin 1/2 particles. We showed how these Green functions can
be obtained by algebraic methods in an extremely compact and elegant
way. We also showed how to reobtain these solutions with an
appropriate ansatz for the Green function and, finally, we discussed
a semiclassical solution and checked that for the case at hand this
approximation is sufficient to give the exact result. From one hand,
this is a surprising result, since the lagrangian involved in the
solution is far from being quadratic. On the other hand, it could
have been guessed, for a semiclassical approximation yields the
exact Green function for a relativistic charged particle in an
external electromagnetic field of a plane wave.
 The results obtained here corroborates and
complements those of reference \cite{Barducci1}. We think that
 our calculations may be useful in other contexts. Particularly, the
 search for other non-quadratic problems whose exact solutions coincide with
 those obtained by the semiclassical approximation is and
 interesting issue and can tell us more about the role of such an
 approximation in physical problems.

\vskip 0.2 cm
 \noindent
 {\bf Acknowledgements:} It is difficult to express in a few words how we are
 indebted to Professor Arvind Narayan Vaidya, who unfortunately is not among us anymore,
  for everything that he has taught us along his academic life. The authors thank to CNPq
  for financial support.


\begin{thebibliography}{99}
%
\bibitem{Barducci3} Barducci A and Giachetti R 2005 {\it J. Phys.} A {\bf 38} 1615.
%
\bibitem{Schwinger1951} Schwinger J 1951 {\it Phys. Rev.} {\bf 82} 664
%
\bibitem{Dodonov1} Dodonov V V, Malkin I A and Man'ko V I 1975
{\it Lett. Nuovo Cimento} {\bf 14} 241
%
\bibitem{Dodonov2} Dodonov V V, Malkin I A and Man'ko V I 1976
{\it Physica} {\bf 82} 113
%
\bibitem{Dodonov3} Dodonov V V, Malkin I A and Man'ko V I 1976
{\it J. Phys.} A {\bf 215} 1791
%
\bibitem{Likken} Likken J D, Sonnenschein J and Weiss N 1991
{\it Int. J. Mod. Phys. A} {\bf 6} 5155
%
\bibitem{Boschi} Boschi-Filho H, Farina C and Vaidya A N 1996
{\it Phys. Lett.  A} {\bf 184} 23
%
\bibitem{VaidyaJPA1988} Vaidya A N, Hott M B and Farina C 1988
{\it J. Phys. A} {\bf 21} 2239
%
\bibitem{Schwinger1964} Schwinger J 1964 {\it J. Math. Phys.}
{\bf 5} 1606
%
\bibitem{Strakhovenko} Milshtein A I and Strakhovenko V M 1982
{\it Phys. Lett. A} {\bf 90} 447
%
\bibitem{ItzyksonZuber}  Itzikson C and Zuber J B 1980
{\it Quantum Field Theory} (McGraw-Hill)
%
\bibitem{Gitman} Fradkin E S, Gitman D M and Shvartsman S H M 1991
{\it Quantum Electrodynamics: with unstable vacuum} (Springer Series
in Nuclear and Particle Physics)
%
\bibitem{Barducci1} Barducci A and Giachetti R 2003 {\it J. Phys.} A {\bf 36} 8129
%
\bibitem{Barducci2} Barducci A and Giachetti R 1976
{\it Lett. Nuovo Cimento} {\bf 15} 309
%
\bibitem{VaidyaNuovoCimento1988}  Vaidya A N, Farina C 1988 {\it Nuovo Cimmento} {\bf 69} 101
%
\bibitem{Volkov1935} Volkov D M 1935  {\it Zeits. Phys.}
{\bf 94} 25
%
\bibitem{BaroneBoschiFarinaAJP} Barone F, Boschi-Filho H and Farina C
 2003 {\it Am. J. Phys.} {\bf 71} 483
%
\bibitem{LivroMecQuantSchwinger} Schwinger J 2001 {\it Quantum Mechanics:
Symbolism of Atomic Measurements} (Springer, ed. Englert B G)
%
\bibitem{FarinaAntonioPLA1993} Farina C and Seguí-Santonja A J 1993
{\it Phys. Lett. A} {\bf 184} 23
%
\bibitem{HoringPRA1986} Horing N J M, Cui H L and Fiorenza G 1986
 {\it Phys. Lett. A} {\bf 34} 612
 %
 \bibitem{Goldberger} Goldberger M {\it Lecture notes on quantum mechanics} (Princeton)
%
\end{thebibliography}
\end{document}